\documentclass{article}

     \PassOptionsToPackage{numbers, compress}{natbib}

     \usepackage[preprint]{neurips_2020}

\usepackage[utf8]{inputenc} 
\usepackage[T1]{fontenc}    
\usepackage{hyperref}       
\usepackage{url}            
\usepackage{booktabs}       
\usepackage{amsfonts}       
\usepackage{nicefrac}       
\usepackage{microtype}      
\usepackage{times}
\usepackage{epsfig}
\usepackage{graphicx}
\usepackage{amsmath}
\usepackage{amssymb}
\usepackage{color,soul}
\usepackage{listings}
\usepackage{multicol}
\usepackage{multirow}
\usepackage[flushleft]{threeparttable}
\usepackage{hyphenat}
\usepackage{subcaption}
\usepackage{varwidth}
\title{Woodpecker-DL: Accelerating Deep Neural Networks via Hardware-Aware Multifaceted Optimizations}

\author{%
	Yongchao Liu\thanks{Equal contributions.}, Yue Jin~\footnotemark[1], Yong Chen, Teng Teng, Hang Ou,  Rui Zhao, Yao Zhang\\
	Ant Group, China \\	
  \texttt{$\left \{\begin{varwidth}{8cm}\centering yongchao.ly, jinyue.jy, cy119846, teng.t, ouhang.oh, rui.rz, yao.zhang\end{varwidth}\right\}@antgroup.com$} \\
}

\begin{document}

\maketitle

\begin{abstract}
Accelerating deep model training and inference is crucial in practice. Existing deep learning frameworks usually concentrate on optimizing training speed and pay fewer attentions to inference-specific optimizations.
Actually, model inference differs from training in terms of computation, e.g. parameters are refreshed each gradient update step during training, but kept invariant during inference. These special characteristics of model inference open new opportunities for its optimization. In this paper, we propose a hardware-aware optimization framework, namely Woodpecker-DL (WPK), to accelerate inference by taking advantage of multiple joint optimizations from the perspectives of graph optimization, automated searches, domain-specific language (DSL) compiler techniques and system-level exploration. In WPK, we investigated two new automated search approaches based on genetic algorithm and reinforcement learning, respectively, to hunt the best operator code configurations targeting specific hardware. A customized DSL compiler is further attached to these search algorithms to generate efficient codes. To create an optimized inference plan, WPK systematically explores high-speed operator implementations from third-party libraries besides our automatically generated codes and singles out the best implementation per operator for use. Extensive experiments demonstrated that on a Tesla P100 GPU, we can achieve the maximum speedup of 5.40 over cuDNN and 1.63 over TVM on individual convolution operators, and run up to 1.18 times faster than TensorRT for end-to-end model inference.
\end{abstract}
\section{Introduction}
\label{section:introduction}
Convolutional neural network (CNN) models~\cite{krizhevsky2012imagenet, ren2015faster,lin2018tsm, he2017mask} usually have high computational cost subject to batch size, number of weight parameters and image size. Hence, graphics processing units (GPUs) have been playing a central role in CNN model training and inference, owing to its high compute power exposed by massive parallelism. Popular deep learning frameworks such as Caffe~\cite{jia2014caffe}, TensorFlow~\cite{abadi2016tensorflow}, Mxnet~\cite{chen2015mxnet} and PyTorch~\cite{paszke2019pytorch} all provide built-in support for GPUs. However, these frameworks mainly focus on improving programming productivity and training performance. Under such circumstances, a few model-inference optimization works have been proposed. They generally work as follows: ($i$) taking as input a trained model from the aforementioned deep learning frameworks, and ($ii$) generating an optimized implementation for deployment in production. Typical works include XLA~\cite{google2019xla} (applicable to training as well), TVM~\cite{chen2018tvm}, Glow~\cite{rotem2018glow}, Tensor Comprehensions~\cite{vasilache2018tensor}, nGraph~\cite{cyphers2018intel}, OpenVINO~\cite{intel2019openvino}, and TensorRT~\cite{nvidia2019tensort}.

The architecture of a deep neural network (DNN) can be abstracted as a computational graph with operators as nodes and tensors representing data movement as edges. In practice, computation within operators often dominates the whole execution, in contrast to data movement between operators. In this case, faster execution of individual operators would lead to prominent acceleration of model inference. Therefore, hardware vendors devote considerable efforts to manually tuning the performance of key primitive functions that are widely used by deep learning applications. These primitives are commonly offered as a collection of libraries, allowing for practitioners to leverage the latest architectural features and refinement in primitive implementations. Existing deep learning frameworks heavily rely on these highly engineered libraries such as \mbox{MKL-DNN}~\cite{intel2019mkldnn} on CPUs and cuDNN~\cite{chetlur2014cudnn} on GPUs. Although these libraries are usually very efficient, there may still be significant room for performance improvement. This is because given a specific primitive, manual tuning is usually unable to explore the whole optimization space, thus possibly missing better implementations. Moreover, these libraries do not implement the full set of primitives needed by deep learning models, leaving the implementation and optimization of those unsupported ones to users. As a matter of fact, implementing high-performance novel primitives is essentially challenging even for experts. This motivated the development of domain-specific language (DSL) compilers to lower programming barrier and thereby allow for non-expert users to write high-performance primitives with no need of deep knowledge of hardware and associated parallel programming models. Typical DSL compilers include Halide~\cite{ragan2013halide}, DLVM~\cite{wei2017dlvm}, Diesel~\cite{elango2018diesel}, TIRAMISU~\cite{baghdadi2019tiramisu} and Triton~\cite{tillet2019triton}.

In principle, accelerating primitives intends to optimize the inference at the operator level. However, after examining popular deep learning models, existing works further observed optimization opportunities from patterned subgraphs that allow for fusing consecutive operators to reduce or even eliminate data movement between the operators fused. Note that implementing a fused operator by invoking primitive functions of component operators one after another is actually unable to reduce data movement overhead between operator calls, making operator fusion ineffective at all. Taking GPU as an example, one effective approach is to write one CUDA~\cite{lindholm2008nvidia} kernel function for the fused operator and complete the whole computation within only one kernel launch to eliminate the intermediate data movement overhead mentioned above. The benefits gained from this in-placed implementation inspired us to perform global optimizations at the graph level. Typical optimizations include fusing operators, removing redundant operations (e.g. identify and dropout), functionally equivalent subgraph substitution~\cite{jia2019taso} and \textit{etc}. Similar to writing  primitives unsupported by vendor-specific libraries, implementing efficient kernels for fused operators is also challenging to users. One promising approach is to resort to DSL compilers.

Have examined existing works on model inference acceleration, we observed that none of them has ever made attempts on leveraging system-level exploration to identify best-performing operator functions additionally from third-party implementations. In this situation, we propose Woodpecker-DL (WPK), a hardware-aware optimization framework that leverages multifaceted optimizations based on local and global graph optimization, automated searches with genetic algorithm~\cite{holland1992adaptation} and reinforcement learning(RL)~\cite{schulman2017proximal}, automatic high-quality code generation by a customized DSL compiler, and system-level exploration to exploit third-party superiorities. Our contributions can be summarized from the following two aspects. On one hand, we proposed an automated optimization framework for model inference acceleration by taking advantage of an ensemble of systematic optimizations coming from computational graph, automated searches, DSL compilers and third-party libraries. This framework allows for non-expert users to achieve high-speed model inference with no need of deep understanding of the underlying hardware architectures and parallel programming models. On the other hand, we developed an automated hardware-aware search method based on RL, named \mbox{RL-search}, besides genetic algorithm. These automated searches free users from the tedious and laborious exploration of the vast search spaces exposed by device-specific primitives.
\section{Methods}
\begin{figure*}[t!]
\centering
\begin{subfigure}[t]{0.49\linewidth}
\includegraphics[width=0.95\linewidth]{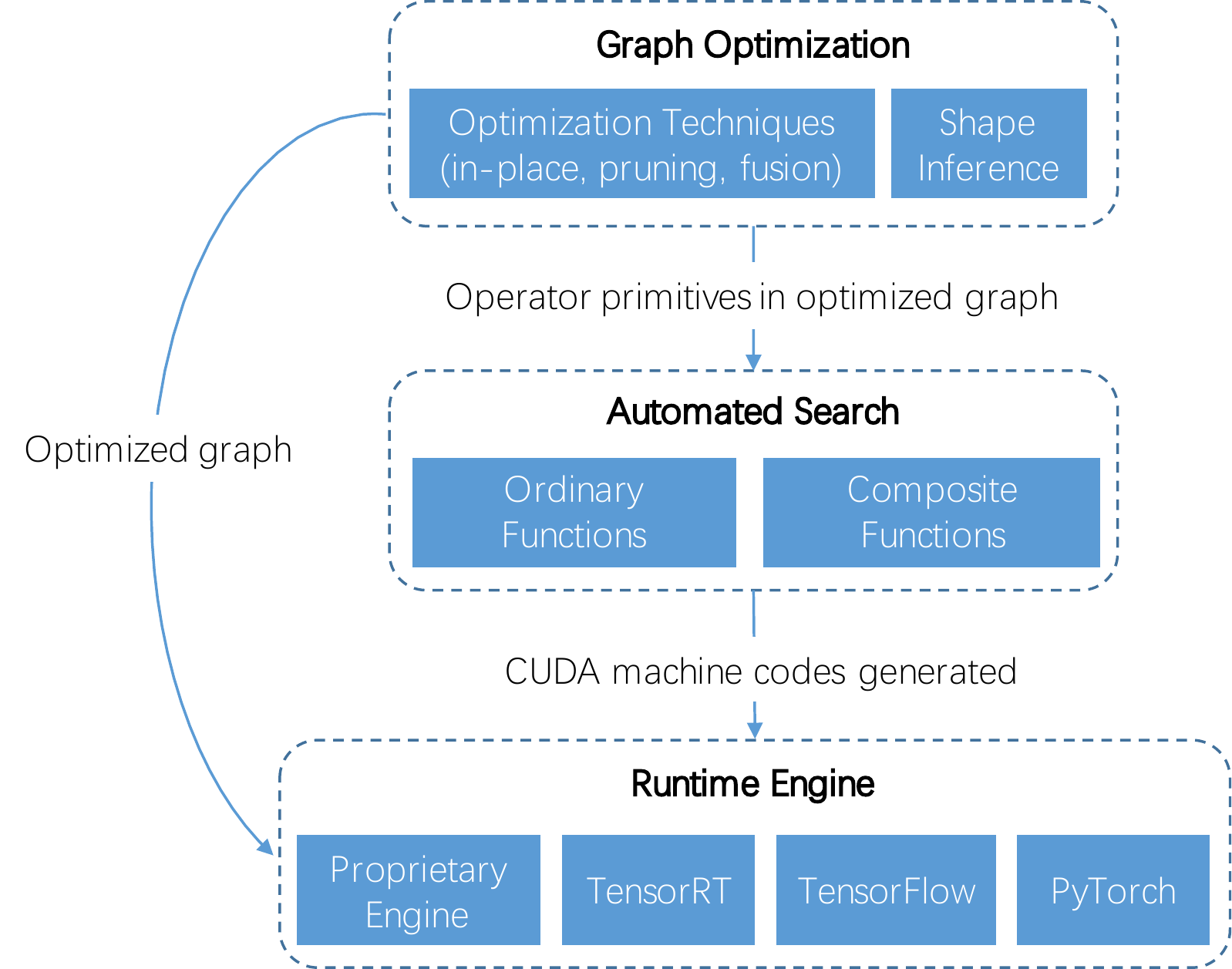}
\caption{}
\label{figure:overview}
\end{subfigure}
%
\begin{subfigure}[t]{0.49\linewidth}
\includegraphics[width=0.95\linewidth]{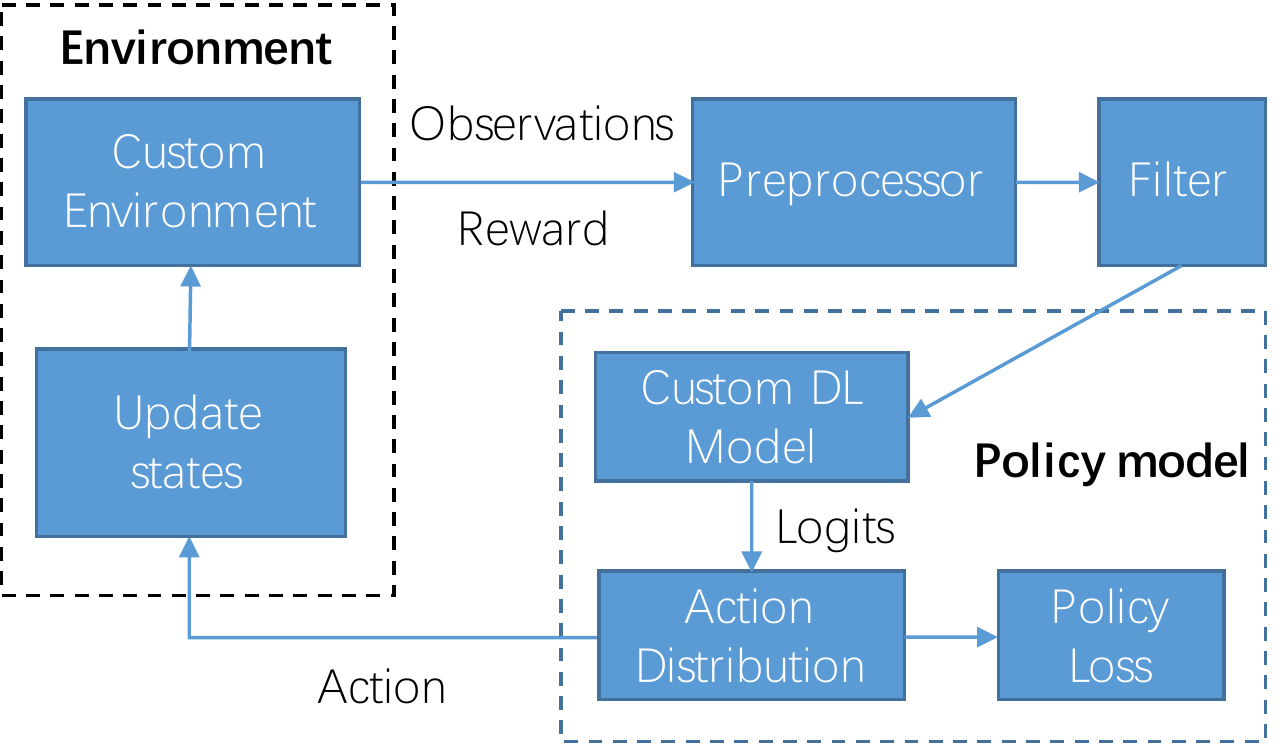}
\caption{}
\label{figure:rl}
\end{subfigure}
\caption{Architectural overview of (a) WPK, and (b) RL-search}
\end{figure*}

In principle, WPK consists of four components: graph optimization, automated search, runtime engine, and custom operators bound to third-party engines. The graph optimization component takes a graph model as input, then performs functionally equivalent transformations to simplify graph structures, and finally outputs the specification that guides automatic code generation per operator in the optimized graph. The automated search component accepts the specifications exported by the graph optimization component and couples genetic search and \mbox{RL-search} with our customized Halide compiler to generate efficient codes for each operator. Our runtime engine collects the operator functions generated by automated searches, and drives the data flow expressed by the optimized graph to complete inference. In addition to our proprietary runtime engine, WPK allows for encapsulating our generated operator functions into custom operators that comply with the standards defined by existing deep learning frameworks (e.g. TensorRT, TensorFlow and PyTorch). Figure \ref{figure:overview} shows the architectural overview of WPK.
\subsection{Graph optimization}
Computational graph optimization has become a standard procedure in accelerating deep neural networks. Most approaches apply pre-defined rules to identify sub-graphs that can be equivalently simplified and substituted. In WPK, we have used the following approaches: constant folding, operator fusion and data layout transformations. Constant folding applies to sub-graphs whose output values can be computed statically beforehand. Operator fusion aims to compress the computation with a sub-graph into one equivalent novel operator in order to reduce the communication overhead between operators in the sub-graph as well as improve hardware usage efficiency due to the increase of compute intensiveness within the novel operator. Data layout transformations aims to identify the better data layouts for the inputs to a given operator in order to get faster execution on the target hardware. It needs to be stressed that for operator fusion and data layout transformations, we must compose the corresponding implementations of those novel operators newly created in the optimization process. In WPK, we employed rule-based optimizations and defined pattern-based specifications to guide the generation of operator functions accordingly.
\subsection{Automated Searches}
Our automated searches intend to identity most efficient codes per operator according to code-generation specifications tailored for a specific architecture. Instead of merely searching hand-engineered libraries, we choose to take advantage of a customized Halide compiler to generate codes just-in-time under the guidance of our search algorithms. Since Halide can support a wide spectrum of processors including x86/ARM CPUs and GPUs, WPK naturally supports these architectures as well. Nonetheless, our paper will merely investigate optimization techniques on CUDA-enabled GPUs~\cite{lindholm2008nvidia}.
\paragraph{Halide compiler}
Halide is a DSL compiler based on the concept of functional programming. A Halide program is actually C++ code written using the functions (\texttt{Func} type), variables (\texttt{Var} type) and other types (e.g. \texttt{Expr}) defined in Halide library. These functions and expression definitions are embedded in C++ syntax by means of operator overloading on the corresponding types. Halide is defined based on the concept of separating algorithm from schedule, where \texttt{algorithm} declares what to compute and \texttt{schedule} represents the decisions about how to map and run the algorithm efficiently on a target device. The following code snippet gives a simple convolution implementation in Halide C++ syntax.
\begin{center}
\begin{lstlisting}[basicstyle=\fontsize{7}{7.5}\selectfont\ttfamily]
// Express the algorithm.
Buffer<float> in(256, 256);
Buffer<float> filter(3, 3);
RDom r(filter);
Func conv;
conv(x, y) = sum(filter(r.x, r.y) * in(x+r.x-1, y+r.y-1));

// Schedule with tunable parameters x_size and y_size;
Var xi, yi;
conv.gpu_tile(x, y, xi, yi, x_size, y_size);

// Generate codes and execute them on the target.
Target target = get_gpu_target_from_environment();
Buffer<float> out = conv.realize(256, 256, target);
\end{lstlisting}
\end{center}
\paragraph{Generating codes}
As mentioned above, we employ Halide to generate codes for operator functions. Intuitively, given an algorithm description, we could make attempts to find an optimal schedule by enumerating all possible configurations in the whole schedule space exposed by Halide. However, this approach will incur huge computation and could result in prohibitively long runtimes, thus inapplicable to practical use. In our implementation, we adopted a semi-automatic approach based on schedule templates. This approach pre-defines one or more schedule templates for a given algorithm, then exposes a set of tunable hyper-parameters to let practitioners instantiate, and finally exploits automated search in the tunable parameter space to identify specific parameter values that are capable of directing optimal code generation. For instance, \texttt{x\_size} and \texttt{y\_size} are tunable parameters in the code snippet shown above. Moreover, due to its confined search spaces, this semi-automatic approach is obviously advantageous to whole-space search approaches in terms of speed. In our implementation, schedule templates are all composed by domain-specific experts, and are fed into Halide at the runtime to generate codes with the assistance of our automated searches.
\subsection{Genetic search}
\label{section:genetic_search}
Genetic algorithms are a family of meta-heuristic optimization algorithms inspired by the principles of natural selection and genetics. These algorithms mimic evolutionary processes by performing crossover, mutation and selection operations. In practice, they are capable of advancing, with high robustness, to optimal solutions to complicated optimization problems. 

\paragraph{Search space}
In WPK, genetic search is used to identify an optimal configuration for code generation per operator on the target hardware. A configuration is encoded as a parameterized vector (or \texttt{chromosome} in the parlance of genetics) $s=\{c_0, c_1, ..., c_{n-1}\}$ of $n$ elements with each element $c_i$ (\mbox{$0\le i < n$}) corresponding to a numerical parameter (or \texttt{gene} with respect to chromosomes) having a finite range. The full set of all configurations $\{s\}$ constitute the search space $S$ of our genetic search. A configuration is hardware-dependent in some sense and is used to instantiate a schedule template.
\paragraph{Implementation}
Our implementation follows the typical procedure of genetic algorithms, which generally consists of four steps: ($i$) \texttt{Step1} initializes a population $a$ of $|a|$ configurations that are randomly generated, ($ii$) \texttt{Step2} calculates the fitness value for each configuration in the population, ($iii$) \texttt{Step3} performs genetic operations including crossover, mutation and selection, and ($iv$) \texttt{Step4} repeats \texttt{Step2} and \texttt{Step3} until the convergence condition is met.
In \texttt{Step1}, any randomly generated configuration will be verified first in order to meet certain constraints. For instance, the total number of threads in a thread block cannot exceed 1024 on a CUDA-enabled GPU. In this case, we must ensure that the product of all dimension values is positive and $\le 1024$ for a thread block. In \texttt{Step2}, for each individual $a_i$, we first compile the generated codes just-in-time as per the hardware configuration, then execute them to get the runtime, and finally set the function of runtime, denoted as \mbox{$f(a_i)$}, as its fitness value.

\texttt{Step3} first calculates the selection probability \mbox{$p(a_i)$} (refer to Equation (\ref{equation:selection})) for \mbox{$a_i$}, and sorts the population in decreasing order of selection probability.
\begin{equation}
p(a_i) = \frac{f(a_i)}{\sum_{i=1}^{|a|}f(a_i)}
\label{equation:selection}
\end{equation} 
Subsequently, we select top $k$ (\mbox{$1\le k \le |a|$}) elites with the highest probabilities. These elites are always selected and passed to the next generation. In addition to these elites, we will further reproduce some off-springs from individuals with less fitness in order for more exploration. Assuming the expected next-generation population size is $|a'|$ (\mbox{$|a'|\ge k$}), we employ a roulette wheel selection approach to randomly select parents for any of the remaining $|a'|-k$ children and crossover to reproduce off-springs. This selection first computes the cumulative probabilities from the selection probabilities of the $m$ (\mbox{$m \le |a|$}) individuals that will participate in the crossover. In this case, the cumulative probability $P(a_i)$ (\mbox{$1\le i \le m$}) of the $i$-th individual is calculated by Equation (\ref{equation:roulette}).
\begin{equation}
P(a_i) = \sum_{j=1}^{i}p(a_j)
\label{equation:roulette}
\end{equation}
Based on this equation, we used an inverse sampling approach to select candidates. More specifically, after getting $P(a_i)$, we generate a random number $v$, which is uniformly distributed in [0, 1], and compare $v$ with $P(a_i)$ to select individuals. If \mbox{$P(a_{i-1}) < v \le P(a_{i})$}, the $i$-th individual will be selected. In sum, the core idea of our selection is to make more healthy individuals breed more and less healthy ones to breed fewer or even nothing.

\texttt{Step4} will stop the evolutionary process as long as the convergence condition is reached, i.e. the runtimes of all individuals in the current generation are close enough. Additionally, note that the population size from generation to generation may vary in our implementation. 
\subsection{Reinforcement learning search}
Besides genetic search described in ~\ref{section:genetic_search}, we have developed RL-search, an automated search algorithm based on RL. Given an operator, we model schedule template parameter optimization as a RL problem, and adopt the proximal policy optimization (PPO)~\cite{schulman2017proximal} approach to predict one action only, i.e. instantiating the schedule template with a concrete parameter configuration. PPO is a new family of policy gradient methods for RL. Unlike standard policy gradient methods ~\cite{mnih2016asynchronous} performing one gradient update per data sample, PPO enables training with mini-batch updates. In addition, RLlib~\cite{liang2017rllib} is used to implement our search algorithm (see Figure~\ref{figure:rl}).
\paragraph{State space}
We introduced a feature vector $O$ to represent our observation, where all possible values of $O$ form our state space. For different operators, we could use distinct observation representation. For 2D convolutions that are basically most time-consuming in CNN models ~\cite{li2016performance}, the observation $O_{conv}$ is 17-dimensional and defined as
\begin{align*}
\begin{array}{ll}
O_{conv}=& (N, C_{in}, C_{out}, K_h, K_w, H, W, Stride, Padding, \\
        & T_x, T_y, T_z, Tile_x, Tile_y, Tile_z, Tile_{rz}, \alpha_t)
\end{array}
\end{align*}
where $N$ is the batch size, $C_{in}$ (and $C_{out}$) is the number of input (and output) channels, $K_h$ (and $K_w$) is the number of rows (and columns) in a filter matrix, $H$ (and $W$) is the image height (and width), $Stride$ is the stride and $Padding$ is the padding mode (i.e. \texttt{SAME} or \texttt{VALID} in our case). $T_x$, $T_y$, and $T_z$ denote the number of CUDA threads in the $x$, $y$ and $z$ coordinate direction of a thread block, respectively, while $Tile_x$, $Tile_y$ and $Tile_z$ are the tile sizes that will be processed by a single CUDA thread in the $x$, $y$ and $z$ coordinate direction, respectively. $Tile_{rz}$ represents the split and unroll size in a reduce domain, $\alpha_t$ (\mbox{$t\ge 1$}) is the runtime moving average of the operator at time step $t$. In our implementation, $\alpha_t$ is empirically calculated as follows:
\begin{equation}
\alpha_t = \frac{\alpha_{t-1} \times 0.8 + \beta_t}{t}
\end{equation}
where $\beta_t$ denotes the runtime of the operator at time step $t$, and $\alpha_0$ is initialized to be zero.
\paragraph{Action space}
We used a discrete action space and developed a DNN to predict actions from observations. The DNN is composed of four fully-connected (FC) layers (with 512, 1024, 1024 and 512 hidden sizes in order) associated with an activation function (\texttt{tahn}, \texttt{tahn}, \texttt{selu} and \texttt{selu} functions in order) each, followed by a dropout layer with a keep probability of 15\%, and a FC layer with a linear activation. The output of the network is fed into a multinomial distribution to sample actions. The output of the multinomial distribution is used as actions to update the parameter values (e.g. $Tile_x$ in $O_{conv}$) in our state space, where an action updates one parameter at a time and multiple rounds of action predictions are required in order to perform the same number of parameter updates.
\paragraph{Network}
We adopted a model-free method which obtains the runtime of the operator by directly interacting with the target hardware. In our implementation, the reward $r_t$ at time step $t$ is defined as:
\begin{equation}
r_t = \alpha_{t-1} - \min\{\beta_t, 2\alpha_{t-1}\}
\end{equation}
The rationale behind $r_t$ is that if $\beta_t$ is less than the historical moving average $\alpha_{t-1}$, we return a positive reward calculated from the runtime difference, and otherwise, a negative reward. If $\beta_t$ is considerably large, say $\beta_t > 2\alpha_{t-1}$ in our implementation, we will clamp its value to $2\alpha_{t-1}$, resulting in the reward of $-\alpha_{t-1}$.

As mentioned above, our RL agent employs the PPO algorithm, whose computation requires computing an estimator of the policy gradient and plugging the estimator into a stochastic gradient ascent algorithm. We adopted the generalized advantage estimator proposed in \cite{schulman2017proximal}, defining the estimator $\bar{A}_t$ of advantage function at time step $t$ as
\begin{equation}
\bar{A}_t = \delta_t + (\gamma\mu)\delta_{t+1} + \cdots + (\gamma\mu)^{T-t+1}\delta_{T-1}
\end{equation}
where
\begin{equation}
	\delta_t = r_t + \gamma V(s_{t+1}) - V(s_t)
\end{equation}
and $V(s_t)$ is the score returned by a learned state-value function at time step $t$.

Our loss function $L_t(\theta)$ combines the policy surrogate loss $L_t^{clip}(\theta)$ with a value function loss $L_t^{VF}(\theta)$, and is further augmented with the addition of an entropy bonus to ensure sufficient exploration, as done in~\cite{schulman2017proximal}. Therefore, we defined the final loss function as
\begin{equation}
L_t(\theta) = \bar{E}_t[L_t^{clip}(\theta) - c_1L_t^{VF}(\theta) + c_2S[\pi_\theta ](s_t)]
\end{equation}
where $c_1$ and $c_2$ are coefficients and are set to 0.15 and 20 in our implementation, respectively. $S$ denotes an entropy bonus and $L_t^{VF}$ the square-error loss $V_\theta(s_t) - V_t^{target}$. Please refer to RLlib \cite{liang2017rllib} for more implementation details.
\subsection{Integration with TensorRT}
\label{section:trt}
\begin{figure*}[t!]
\centering
\begin{subfigure}[b]{0.55\linewidth}
\includegraphics[width=\linewidth]{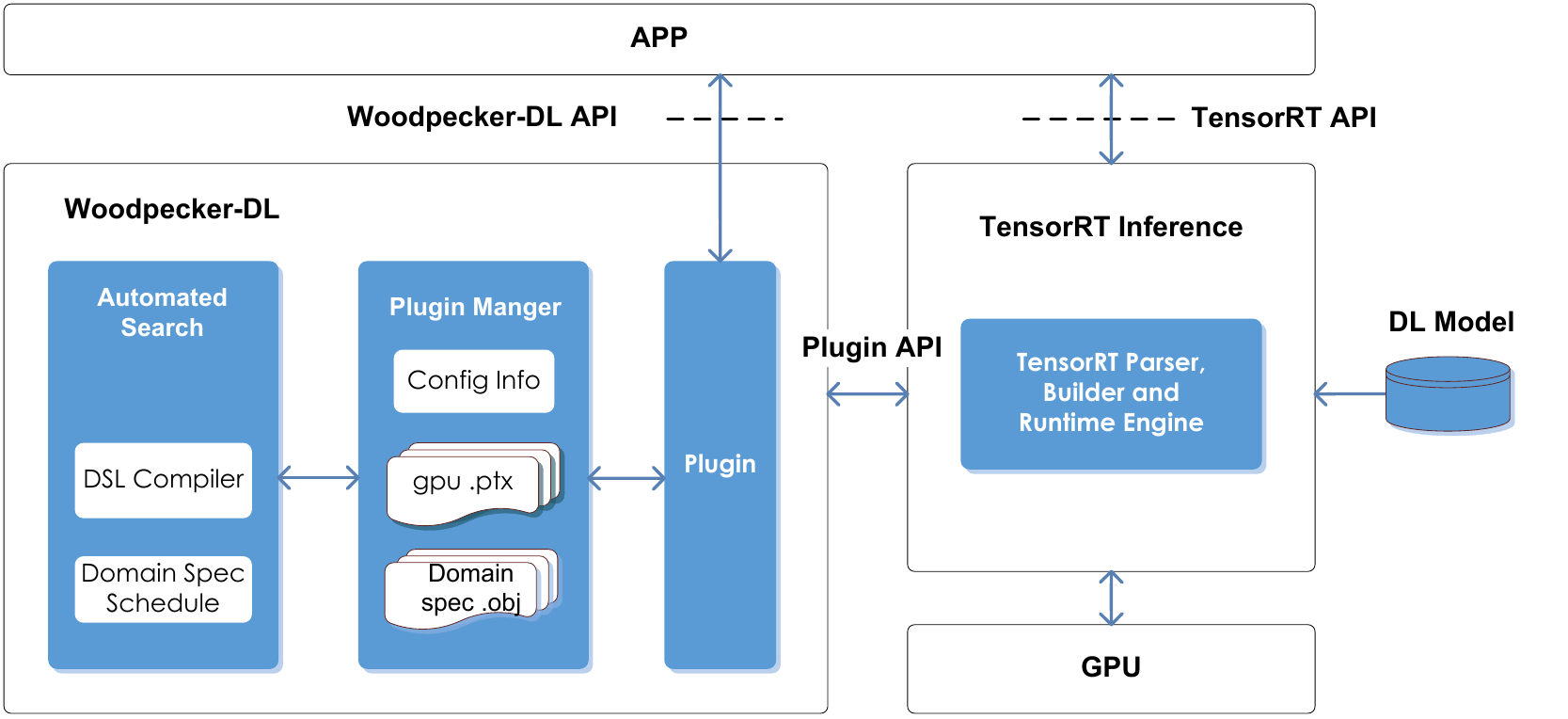}
\caption{}
\label{figure:wpk_trt}
\end{subfigure}
\begin{subfigure}[b]{0.44\linewidth}
\includegraphics[width=\linewidth]{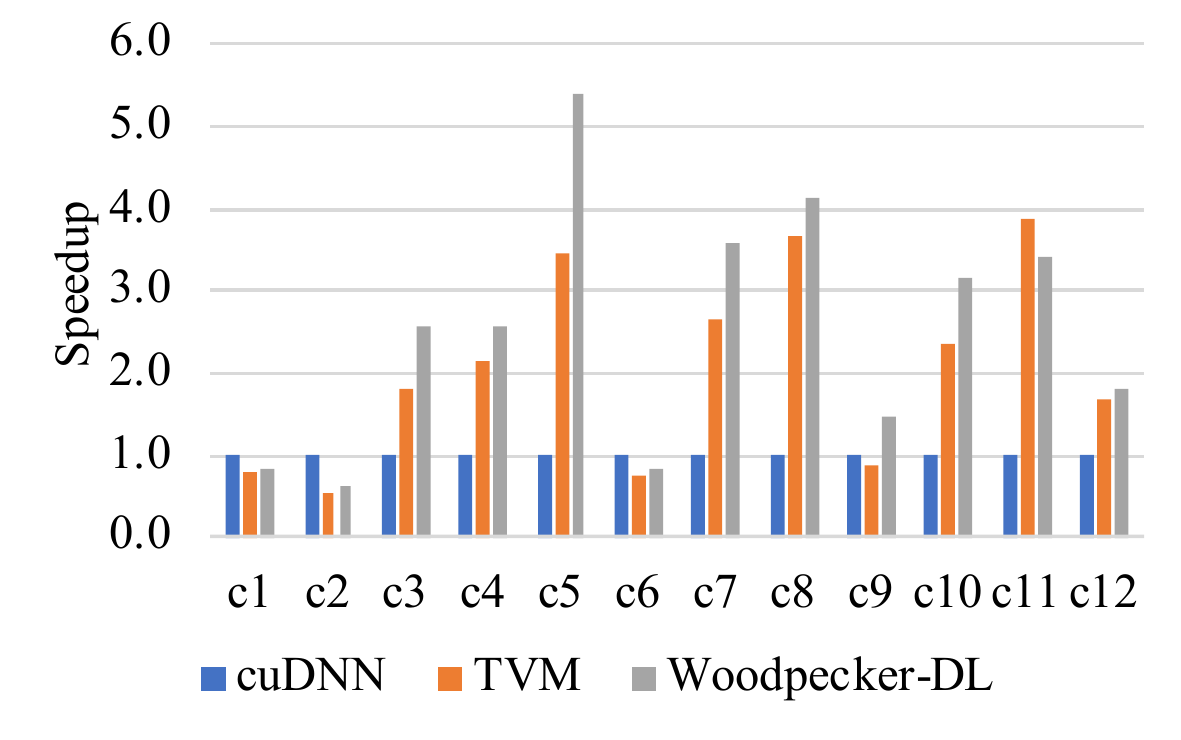}
\caption{}
\label{figure:resnet_ops}
\end{subfigure}
\caption{(a) diagram illustrating the integration of WPK into TensorRT, and (b) speedups of WPK and TVM relative to cuDNN.}
\end{figure*}

TensorRT is a state-of-the-art inference platform on CUDA-enabled GPUs. One special feature of TensorRT is that it allows for users to customize operators via plugins. Based on this feature, we can conveniently integrate WPK-generated codes with TensorRT.
(see Figure~\ref{figure:wpk_trt}).
As mentioned in~\ref{section:introduction}, WPK performs system-level exploration to further use high-performance third-party implementations per operator additionally. This means that we not only take advantage of efficient codes generated by our DSL compiler, but also fully exploit the implementations from third-party libraries (e.g. cuDNN or TensorRT) that outperform ours. Taking TensorRT as an example, for some operator, if its TensorRT implementation is superior to ours, we will use this TensorRT implementation in our optimized inference plan. This type of system-level exploration significantly distinguishes WPK from all existing compiler frameworks including XLA, TVM and nGraph.
\section{Experiments}
We used ResNet-18~\cite{he2016deep} to evaluate WPK and its counterparts on a Tesla P100 GPU. ResNet-18 is an image classification model trained with Caffe and accepts inputs with \texttt{NCHW} data layout format. In terms of end-to-end inference, given an operator, we used both genetic search and RL-search to identify optimal code generation configurations and single out the best for use. WPK was integrated with TensorRT as described in~\ref{section:trt} for inference performance assessment. Additionally, the input shape has $N=1$, $C=3$, $H=224$ and $W=244$.
\subsection{Individual convolution operators}
Firstly, we compared WPK to TVM and cuDNN using the individual convolution operators extracted from ResNet-18. In this test, we categorize convolution operators into distinct groups under the following criterion: two convolution operators are considered computationally identical if they have the same input/output shape, filter matrix size, stride and padding. 
In this test, we directly used the well-optimized \mbox{ResNet-18} model built-in TVM for fair comparison. By using the performance of cuDNN as the baseline, Figure~\ref{figure:resnet_ops} shows the speedups of WPK and TVM relative to cuDNN. From the figure, we can observe that WPK and TVM run $2.54\times$ and $2.06\times$ faster than cuDNN on average, as well as $5.40\times$ (on convolution \texttt{c5}) and $3.89\times$ (on convolution \texttt{c11}) at the maximum, respectively. Interestingly, neither WPK nor TVM is always superior to cuDNN. In comparison with TVM, WPK outperforms the former by a factor of 1.24 on average and 1.63 at the maximum. However, we did not compare with TensorRT, because the overall runtime of TensorRT cannot be broken down as per operator, due to the more complex graph optimizations applied to the model by itself.
\begin{figure*}[t!]
\centering
\begin{subfigure}[b]{0.49\linewidth}
\includegraphics[width=0.9\linewidth]{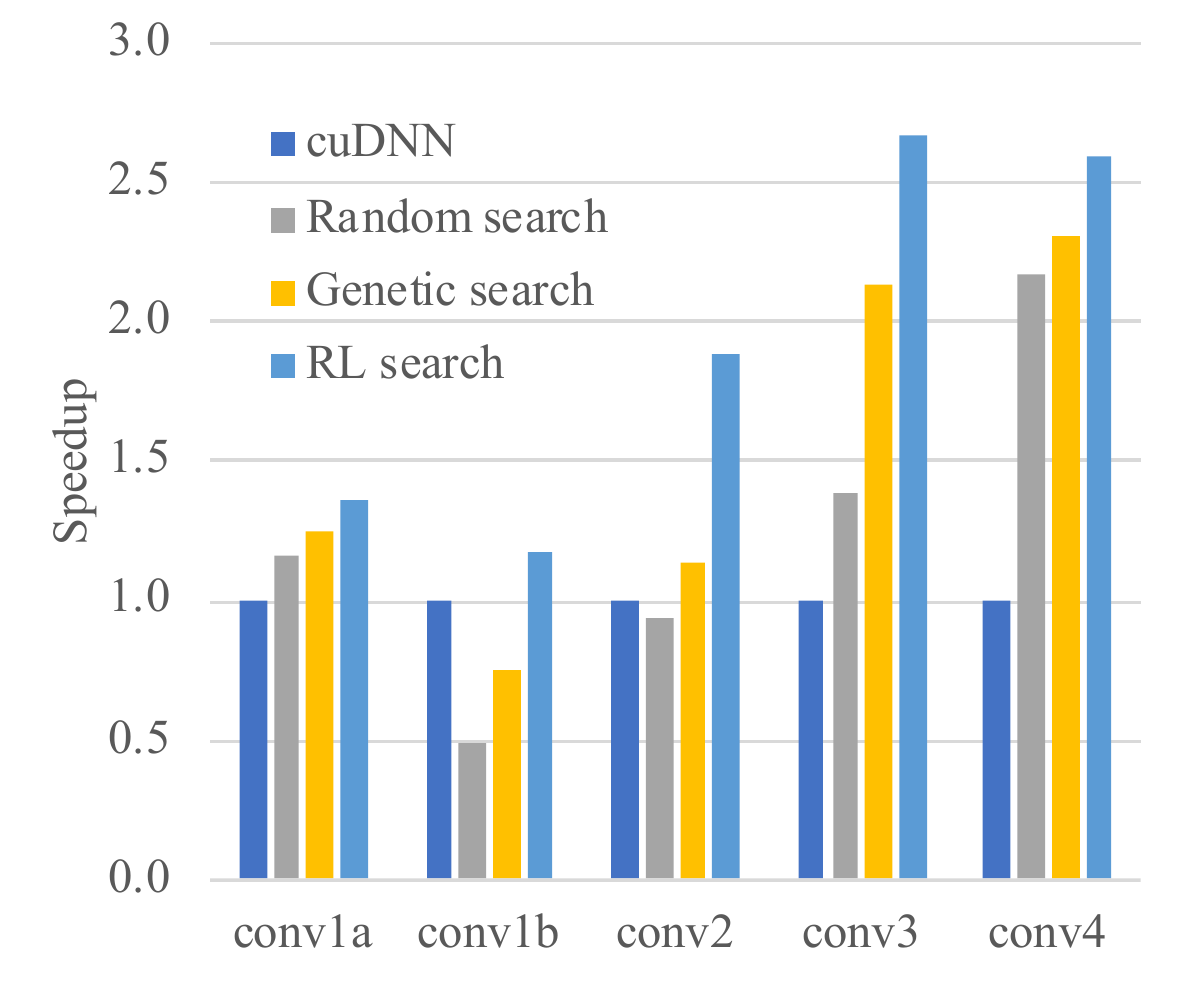}
\caption{}
\label{figure:rl_search}
\end{subfigure}
\begin{subfigure}[b]{0.49\linewidth}
\includegraphics[width=0.9\linewidth]{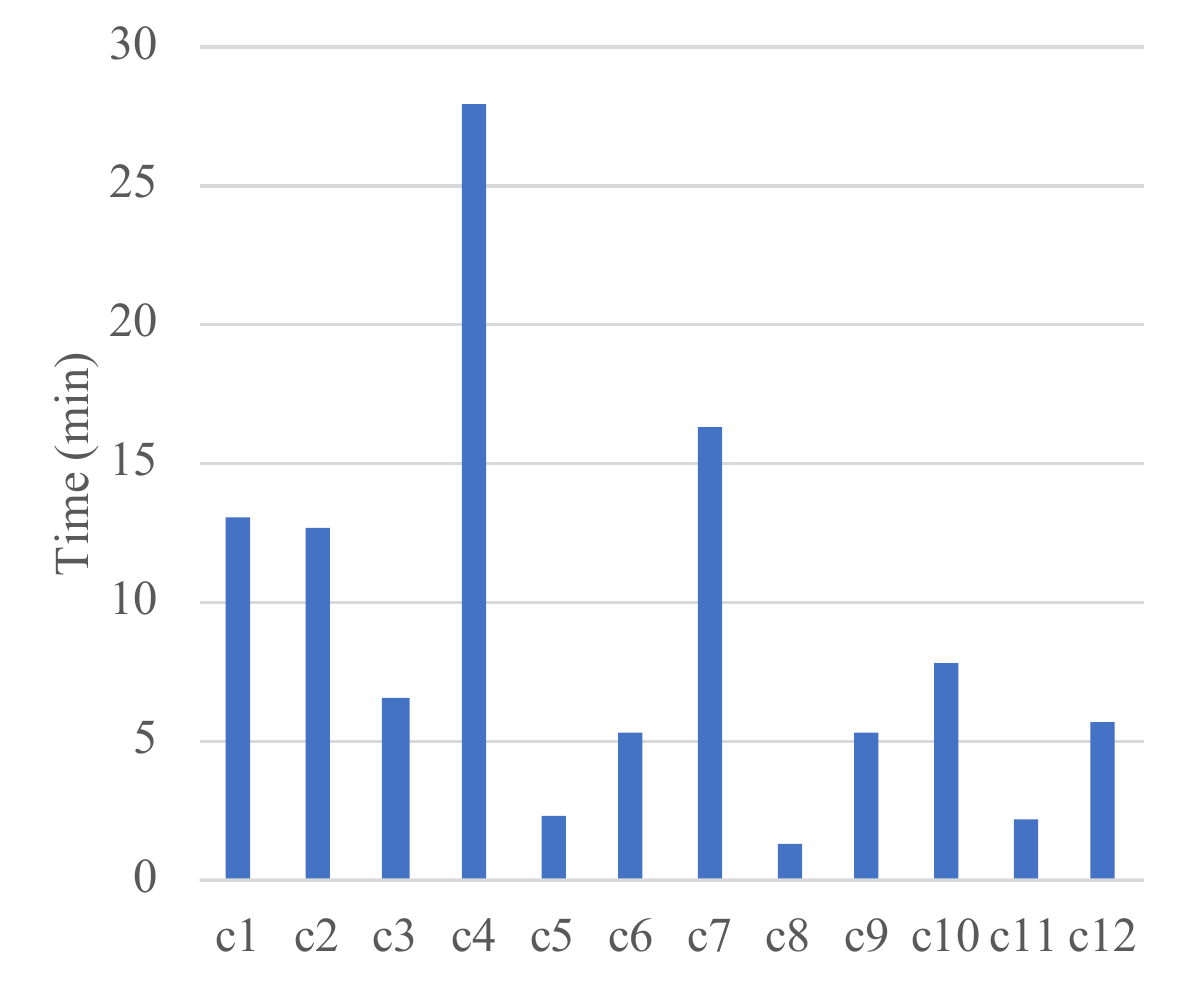}
\caption{}
\label{figure:genetic_search_time}
\end{subfigure}
\caption{(a) performance comparison among three search methods, and (b) genetic search speed on individual convolution operators.}
\end{figure*}
\subsection{RL search performance}
Secondly, we compared RL-search with genetic search. RL-search failed to perform better than genetic search on individual convolutions of ResNet-18. Furthermore, the former was observed to have much higher randomness than the latter, in terms of search time and best operator speed. Nonetheless, we fortunately found that the former yielded superior performance on some convolution operators in another CNN model used in production. Table~\ref{table:rl_convs} gives the information of the convolutions on which RL-search outperforms genetic search, while Figure~\ref{figure:rl_search} shows the speedups of random search, genetic search and RL-search, relative to cuDNN~\cite{chetlur2014cudnn}. From the figure, both RL-search and genetic search consistently outperform random search. In particular, RL-search performs better than genetic search for each case, with speedups ranging from 1.09 to 1.66.
\begin{table}[t]
\caption{Convolutions on which RL-search outperforms genetic search.}
\label{table:rl_convs}
\centering
\begin{tabular}{lllllll}
\toprule
Name &H &W &$C_{in}$ &$C_{out}$ &$K_h\times K_w$ &Stride\\
\midrule
conv1a &112 &96 &3 &64 &$3\times3$ &1 \\
conv1b &110 &94 &64 &96 &$3\times 3$ &2 \\
conv2 &54 &46 &96 &128 &$3\times 3$ &2\\
conv3 &26 &22 &128 &256 &$3\times 3$ &2\\
conv4 &12 &10 &256 &512 &$3\times 3$ &1 \\
\bottomrule
\end{tabular}
\end{table}
\subsection{Genetic search speed}
Thirdly, we evaluated the search speed of our genetic search on individual operators of ResNet-18 (see Figure~\ref{figure:genetic_search_time}). In our implementation, we employed multi-threading to accelerate code compilation as well as generation, and introduced a caching mechanism to reuse search results. The average search time is 8.9 minutes, with the minimum and maximum times of 1.4 and 27.9 minutes respectively. These times are reasonably acceptable in our production, since the search process is normally conducted offline. In addition, our caching mechanism can further expedite the search process for a family of models that are composed from the same backbone model (e.g. ResNet).
\subsection{End-to-end inference}
Finally, we used ResNet-18 to assess the end-to-end inference speed of WPK, TVM and TensorRT. As mentioned in~\ref{section:trt}, WPK can exploit system-level exploration to take advantage of TensorRT operator implementations that run faster than the codes generated by our own compiler. Performance evaluation revealed that WPK is neck-by-neck with TVM, while TensorRT performs worst. WPK runs 1.18$\times$ faster than TensorRT. Note that WPK was observed to have selected some TensorRT operators that outperform WPK-generated codes. Excluding these TensorRT operators incorporated only results in very marginal performance loss of 2\%.
\section{Related Work}
Compiler-based inference acceleration frameworks have been becoming more popular recently. XLA~\cite{google2019xla} is the first work in this research direction, which was initially specialized to TensorFlow models and currently can be applied to optimize PyTorch models as well. XLA lowers operators into primitive linear algebra operations and calls into backend-specific libraries for execution on different backends. TVM~\cite{chen2018tvm} is an end-to-end compiler framework with Halide at the core, which first optimizes a computational graph, then converts the optimized graph into intermediate representations and finally compiles to executable codes on a specific target device. This work was further enhanced by AutoTVM~\cite{chen2018learning} to enable automatic optimization of tensor operators. Compared to TVM, WPK provides broader capability by enabling system-level exploration as described before, i.e. we aim to achieve fastest speed by singling out operator implementations not only from ours but also from third-party libraries. NeoCPU~\cite{liu2019optimizing} is built upon TVM and aims to optimize CNN inference on CPUs by taking advantage of wide SIMD instructions. nGraph~\cite{cyphers2018intel} adopts a similar workflow to TVM, but was further extended to support encypted data with homomorphic encryption \cite{boemer2019ngraph}. Some other compiling frameworks (e.g. Tensor Comprehensions~\cite{vasilache2018tensor}, and Glow~\cite{rotem2018glow}) were also developed.
\section{Conclusion}
WPK is part of Woodpecker that is an efficient compiler framework for heterogeneous computing based on software-hardware co-design, and targets to accelerate deep learning applications by taking advantage of multiple joint optimizations from graph optimization, automated searches, compiling technique and system-level exploration. In this paper, we have presented two automated search methods based on genetic and RL algorithms, respectively. In comparison with cuDNN, TVM and TensorRT, our performance evaluation demonstrated the superiority of WPK in terms of both accelerating individual convolution operators and end-to-end inference. More specifically, on a Tesla P100 GPU, we can achieve the maximum speedup of 5.40 over cuDNN and 1.63 over TVM in terms of individual convolutions, and run up to $1.18\times$ faster than TensorRT with respect to end-to-end model inference. Although we have merely investigated the capability of WPK in accelerating inference in this paper, WPK can actually be applied to accelerate training and we plan to conduct this research as part of our future work. In the end, we would like to note that optimizing device placement of operators in a multi-device environment~\cite{mirhoseini2018hierarchical, narayanan2019pipedream} is also an interesting research direction.
{\small
\bibliographystyle{ieeetr}
\bibliography{woodpecker.bib}
}

\end{document}